\documentstyle[aps]{revtex}
\begin{document}

\twocolumn[\hsize\textwidth\columnwidth\hsize\csname
@twocolumnfalse\endcsname
\title{Asymptotic Behavior of Perturbations in 
Randall-Sundrum Brane-World}
\author{Takahiro Tanaka}
\address{Yukawa Institute for Theoretical Physics, 
Kyoto University, Kyoto 606-8502, Japan}
\maketitle
\begin{abstract}
The asymptotic behavior of metric perturbations in 
Randall-Sundrum infinite brane world is carefully investigated. 
Perturbations generated by matter fields on the brane 
are shown to be regular even at the future Cauchy horizon. \\ 
{YITP-00-32}
\end{abstract}
\vspace{5mm}
]
\section{Introduction}
There are a lot of recent discussions about the possibility 
of existence of extra dimensions in a non-trivial form\cite{gia,RS1,RS2}. 
After the proposal of models with warped 5th-dimension by Randall and
Sundrum\cite{RS1,RS2}, 
the behavior of gravitational perturbations in these models 
has been investigated by many 
authors\cite{CG,CHR,SMS,gt,EHM,CGRT,Rub,SSM,tm,Chiba,GKR}. 
Cosmology based on these models is also 
discussed much\cite{CGRT,Fla,Bin,mukoyama1,Ida,GS,MSM,MWBH,Mukoyama,KIS,Maar,lang,BDBL,KS}. 
The main purpose of such studies will be to find a clue which 
can observationally distinguish these models from 
4-dimensional Einstein gravity. 

As far as the author knows, any manifest contradiction 
with observations has not been reported on these models. 
However, as was pointed out by Randall and Sundrum themselves, 
there has been a worry about the model with infinite 
warped 5-th-dimension from the beginning.
Usually, 5-dimensional Gravitational perturbations are analyzed by 
decomposing them into eigen modes of 4-dimensional d'Alembertian, 
whose eigen values are refered to as comoving mass squared.  
As long as we analyze the behavior of perturbations mode by mode,  
the non-linear interaction seems to become strong as we move far away 
from the brane\cite{RS2}. 
{}Furthermore, by discussing the behavior of massless modes, 
the appearance of instability at the future Cauchy horizon was suggested 
by Chamblin and Gibbons\cite{CG}.  
Appearance of such a bad behavior seemed to be confirmed by the study of 
the asymptotic behavior of the Green's function by Sasaki, 
Shiromizu and Maeda\cite{SSM}. 

If this bad behavior is real, 
is it a serious drawback of the model? 
One way of excuse is ready by Randall and 
Sundrum\cite{RS2}. ``The analysis based on effective theory 
will not be valid there, and instead the more fundamental 
string theory will regulate the situation.'' 
Although this excuse is hard to be denied,  
it is also true that predictability will be reduced in such 
a model which contains spacetime region where unknown 
physics governs. 

Alternatively, one may claim as follows. 
``In realistic cosmological models as discussed in 
Refs.\cite{Fla,Bin,mukoyama1,Ida,GS,MSM,MWBH}, 
we can consider models in which 
future Cauchy horizon does not exist, and hence we are free from 
instability.'' However, the metric in cosmological models at a 
late epoch is not so largely different from the original 
flat model. Hence, if a singularity is developed in the 
original flat model, a rather extreme phenomena will also happen 
in cosmological models even though it might be regularized so as 
not to develop into a singularity. Hence, we will not be able to 
say that the brane world is safe enough for us to live in. 

One may think that another excuse is possible. 
``Even though such a bad behavior will develop at the place far 
away from our brane, we can expect 
that this information will not propagate to our brane, 
and hence it will not cause any disaster 
on the dynamics of our brane world. 
Since in the flat-brane model the singularity develops 
at the future Cauchy horizon even if it exists, it 
will not be seen by the observer living on the brane. 
Therefore, this singularity will be harmless.''
We will see that this kind of excuse is too naive. 
As we have shown in Appendix, 
we can obtain a model in which the brane crosses 
the ``future Cauchy horizon'' by changing the dynamics 
of the brane only in the sufficiently distant future.  

In the present paper, we would like to pursue another 
more attractive possibility which was suggested in 
Ref.\cite{GKR} (See also\cite{RS2}). 
In the paper by Garriga and Tanaka\cite{gt}, it was shown 
that the asymptotic behavior of the 5-dimensional 
gravitational field in static configurations 
is regular at the level of liner perturbation. 
In this calculation, we have seen a non-trivial 
cancellation between contribution from massless modes and 
that from massive K-K modes. 
Hence, we would be able to expect that in general the asymptotic 
behavior of perturbations is much milder than was anticipated 
naively. 
Actually what we shall show in the present 
paper will be that a similar 
cancellation occurs even when the source of 
gravitational perturbations on the brane is dynamical. 

This paper is organized as follows. 
In section 2 the model of Randall-Sundrum brane world 
with infinite extra-dimension is explained. 
Also we give a brief review of the formulation 
for the analysis of 5-dimensional metric perturbations given by
Ref.\cite{gt}.  
In section 3 the asymptotic behavior of 
5-dimensional metric perturbations near the future 
Cauchy horizon is discussed. 
Then, in section 4 we show that 
no singular behavior in metric perturbations 
develops near the future Cauchy horizon. 
Section 5 is devoted to discussion.  

\section{Model}
In this section, we briefly explain the model proposed by Randall and 
Sundrum\cite{RS2}, and review the formulation of evaluating 
metric perturbations induced by matter fields which are confined on the 
brane developed by Ref.\cite{gt}.  
The model of background metric is given by 5-dimensional AdS space 
\begin{equation}
 ds^2= dy^2 + e^{-2|y|/\ell} (-dt^2 + d{\bf x}^2), 
\end{equation}
with a single positive tension 3-brane located at $y=0$. 
Here $\ell$ is the curvature radius of 5-dimensional AdS space. 
We assume reflection symmetry at $y=0$. 

In the Randall-Sundrum gauge, in which $y$-components 
of metric perturbations are set to be zero, 
the equation which governs transverse and traceless 
perturbations of bulk metric tensor 
$h_{\mu\nu}$ induced by matter fields on 
the brane is 
\begin{eqnarray}
&&\left[\ell^{-2} e^{2|y|/\ell}\Box^{(4)} +\partial_y^2 -4\ell^{-2} 
  +4\ell^{-1}\delta(y) \right]h_{\mu\nu} \cr
&&\qquad=-2\kappa\Sigma_{\mu\nu}
  \delta(y), 
\label{heq}
\end{eqnarray}
where $\mu,\nu-$indices run form 0 to 3, and 
$\Box^{(4)}$ is the 4-dimensional flat d'Alembertian. 
Here $\Sigma_{\mu\nu}$ is the source function determined by 
the energy momentum tensor of the matter fields on the brane
$T_{\mu\nu}$ as 
\begin{equation}
 \Sigma_{\mu\nu}=
 \left(T_{\mu\nu}-{1\over 3}\gamma^{\mu\nu} T^\rho{}_\rho\right)
 +2\kappa^{-1} \hat\xi^5_{,\mu\nu}, 
\end{equation}
and $\hat\xi^5$ is determined by solving 
$
 \Box^{(4)}\hat\xi^5 ={\kappa\over 6} T^\rho{}_\rho. 
$
The source function is rather complicated, but what we shall use 
later is just the fact that this tensor is transverse and traceless, 
which is a consequence of conservation law of $T_{\mu\nu}$.
This condition guarantees that induced metric perturbations 
are also transverse and traceless. 

Solving the above equation with the boundary condition 
of no incoming waves from the past Cauchy horizon, 
$y=\infty$ and $t=-\infty$,  
we investigate the asymptotic behavior of $h_{\mu\nu}$ 
near the future Cauchy horizon, $y=\infty$ and $t=+\infty$. 
To solve the above equation, we use the Green's function method. 
Using the retarded Green's function that solves the equation 
\begin{eqnarray}
&&\left[\ell^{-2} e^{2y/\ell}\Box^{(4)} +\partial_y^2 -4\ell^{-2} 
  +4\ell^{-1}\delta(y) \right]G_R (x,x')\cr
 &&\qquad  =\delta^{(5)}(x-x'), 
\end{eqnarray}
the solution of (\ref{heq}) with appropriate boundary condition 
is formally given by 
\begin{equation}
 h_{\mu\nu}(x) = -2\kappa \int d^4 x'\, G_{R}(x,x')\Sigma_{\mu\nu}(x'). 
\label{hmunu}
\end{equation}

The Green's function is constructed by taking a sum over 
a complete set of eigen states 
as usual. Following \cite{RS2,gt}, we have 
\begin{eqnarray}
 G_{R}(x,x') &= &-\int {d^4 k\over (2\pi)^4}
   e^{ik_{\mu}(x^{\mu}-x'{}^{\mu})}
   \Biggl[{e^{-2(|y|+|y'|)/\ell}\ell^{-1}\over {\bf k}^2 -(\omega+i\epsilon)^2}
\cr
   &&\qquad+\int_0^{\infty} dm{u_m(y) u_m(y')\over m^2 
    +{\bf k}^2-(\omega+i\epsilon)^2} \Biggr], 
\end{eqnarray}
with 
$ u_m(y)=\sqrt{m\ell/ 2} (J_1(m\ell) Y_2(m z)
   -Y_1(m\ell) J_2(m z)) $ 
$/ \sqrt{J_1(m\ell)^2 +Y_1(m\ell)^2}$, 
where $J$ and $Y$ are Bessel functions. 
Here we have also introduced the conformal coordinate $z:=\ell e^{y/\ell}$. 
In terms of $z$, the background metric is written as 
\begin{equation}
 ds^2={\ell^2\over z^2}\left(dz^2-dt^2+d{\bf x}^2\right),
\end{equation}
and the brane is located at $z=\ell$. 
When we use $z$, we suppose that we are considering the 
region $z\geq\ell$. 

\section{asymptotic behavior of metric perturbations}

In this section, 
we investigate the asymptotic behavior of metric perturbations 
near the future Cauchy horizon in the original Randall-Sundrum model. 
Our assumptions are that the source $\Sigma_{\mu\nu}$ does not 
continue to exist from infinite past, i.e., it differs 
from zero only for $t>t_0$ with a certain fixed value $t_0$, 
and that the source does not extend to the spatial infinity. 

It is convenient to rewrite the Green's function by using formula
\begin{eqnarray}
&& -\int {d^4 k\over (2\pi)^4}
   {e^{ik_{\mu}(x^{\mu}-x'{}^{\mu})}
     \over m^2 
    +{\bf k}^2-(\omega+i\epsilon)^2} \cr
&&\qquad ={1\over 2\pi}\theta(t-t'){d\over d\sigma^2}
   \left[\theta(\sigma^2) J_0(m\sigma)\right], 
\end{eqnarray}
where $\sigma^2:=(t-t')^2 - (x-x')^2$. 
Setting $y'=0$, 
after a little calculation, we arrive at a rather 
concise expression 
\begin{equation}
 G_R(x^a;0,t',{\bf x}'{}^{\mu})
  ={\theta(t-t')\over 4\pi}{1\over\sigma}
   {\partial\over \partial\sigma} \left[\theta(\sigma^2)
   {\cal G}(\sigma,z)\right],  
\end{equation}
with
\begin{equation}
   {\cal G}(\sigma,z)=
 {i\over 4\pi}\int_{-\infty}^{\infty} dm 
    {H_0^{(2)}((m+i\epsilon)\sigma) 
    H_2^{(1)}((m+i\epsilon)z)\over 
    H_1^{(1)}((m+i\epsilon)\ell)},   
\label{calG}
\end{equation}
where $H^{(i)}$ represents the Hankel function and 
$\epsilon$ is a small positive real constant introduced 
to specify the path of integration. 
It will be manifest that ${\cal G}(\sigma,z)=0$ for 
$z-\ell >\sigma$. 

Now we are interested in the behavior at $\sigma, z\to \infty$. 
For clarity, we specify the way how we take this limit. 
For this purpose, 
we introduce 4-dimensional Milne coordinates by 
\begin{equation}
 x^{\mu}=T\hat x^{\mu}(\Omega),
\end{equation}
where $\Omega$ is a set of coordinates on 3-dimensional unit hyperboloid, 
and $\hat x^{\mu}(\Omega)$ defines its embedding in 4-dimensional Minkowski 
space. 
In the following, we fix $\hat x^\mu$ and keep $z=O(\sigma)$
when we take the $\sigma\to \infty$ limit. 

For this limit, 
the Hankel functions in the numerator 
in (\ref{calG}) can be replaced 
with their asymptotic series expansions,  
\begin{eqnarray}
 H_0^{(2)}(m\sigma)&\sim &
    \sqrt{2\over \pi m\sigma} e^{-im\sigma+\pi i/4}
    \sum_{j=0}^{\infty} {A_j\over (m\sigma)^j}, \cr
 H_2^{(1)}(mz)&\sim &
    \sqrt{2\over \pi mz} e^{imz-5\pi i/4}
    \sum_{j=0}^{\infty} {B_j\over (mz)^j},   
\end{eqnarray}
where $A_j$ and $B_j$ are constant coefficients,  
whose explicit form is not necessary for the present purpose. 
Then we find 
\begin{eqnarray}
   {\cal G}(\sigma,z)&\sim&
  {\theta(\sigma-z+\ell)\over 2\pi^2\sqrt{\sigma z}}
\int_{-\infty}^{\infty} dm 
    {e^{im(z-\sigma)} \over 
      H_1^{(1)}((m+i\epsilon)\ell)}
\cr &&\qquad \times
\sum_i\sum_j {A_i B_j\over (m+i\epsilon)^{i+j+1}\sigma^i z^j}. 
\end{eqnarray}
Let us define 
\begin{equation}
F(\sigma,z)={\kappa\over 8\pi^3}
     {\theta(\sigma-z+\ell)\over \sqrt{\sigma z}}\sum_{i,j} 
     f_{ij}(\sigma-z) \sigma^{-i} z^{-j},
\end{equation}
with 
\begin{eqnarray}
  f_{jk}(s) &=&\int_{-\infty}^{\infty} dm 
    {-ie^{-im s} \over 
      H_1^{(1)}((m+i\epsilon)\ell)}
         {A_j B_k\over (m+i\epsilon)^{j+k+1}}\cr
 &=&\int_{0}^{\infty} d\mu 
    {-\pi^2 ie^{-\mu s} I_1^2(\mu\ell)\over 
      K_1^2(\mu\ell)+\pi^2 I_1^2(\mu\ell)}
         {A_j B_k\over (-i\mu)^{j+k+1}}, 
\end{eqnarray}
where $I$ and $K$ are modified Bessel functions. 
Then metric perturbations (\ref{hmunu}) are written as 
\begin{eqnarray}
 h_{\mu\nu} & \sim & 
  \int d^4 x'\, \Sigma_{\mu\nu}{1\over \sigma}{\partial\over \partial{\sigma}}
   F(\sigma,z). 
\label{hform}
\end{eqnarray}
The $\sigma$-derivative of $F(\sigma,z)$ will produce various kind 
of terms, for some of which term-by-term integration over $x'$ is 
not well-defined separately. 
Hence, it is clearer to rewrite the expression by using integration 
by parts as 
\begin{eqnarray}
h_{\mu\nu} & \sim & 
    -\int d^4 x' \Sigma_{\mu\nu} 
  {\hat x^{\rho}\over (\hat x\cdot \Delta x)}\partial'_{\rho} F(\sigma,z)
\cr
 &=& \int d^4 x'F(\sigma,z)
  \left({1\over (\hat x\cdot \Delta x)^2}
   -{\hat x^{\rho}\over (\hat x\cdot \Delta x)}\partial'_{\rho}\right)
     \Sigma_{\mu\nu},\cr
&&  
\label{hform2}
\end{eqnarray}
where $\partial'_\rho$ is an abbreviation of $\partial/\partial
x'{}^\rho$. 
In the above calculation, we have used 
\begin{eqnarray}
{1\over \sigma}{\partial\over \partial{\sigma}} f(\sigma) 
 & = &  {\hat x^\rho\over (\hat x\cdot \Delta x)} 
\partial'_{\rho} f(\sigma), 
\label{form1}
\end{eqnarray}
which is derived from 
\begin{eqnarray}
 \partial'_{\rho} f(\sigma)
 & = & {\Delta x_\rho\over \sigma}{\partial\over \partial{\sigma}} f(\sigma). 
\label{form2}
\end{eqnarray}
The support of integrand 
in (\ref{hform2}) is finite because of our assumption on 
$\Sigma_{\mu\nu}$ and the existence of step function 
in the expression of $F(\sigma,z)$. 
Since $F(\sigma,z)$ is at most of $O(1/\sigma)$ and 
$(\hat x\cdot \Delta x)=O(\sigma)$, we can conclude that 
at most 
$$
h_{\mu\nu}=O(1/\sigma^2).
$$ 
The result obtained so far  
is similar to that obtained in Ref.\cite{SSM}. 

The next step is to show that 
the components contracted with $\hat x^{\mu}$ 
are suffered from further suppression. 
To show this, we use the property that the source tensor  
$\Sigma_{\mu\nu}$ is transverse and traceless. 
The components contracted with $\hat x^{\mu}$ 
\begin{eqnarray}
\hat x^{\mu} h_{\mu\nu} & \sim & 
\int d^4 x'F(\sigma,z)\cr
&&\quad  \times\left({1\over (\hat x\cdot \Delta x)^2}
   -{\hat x^{\rho}\over (\hat x\cdot \Delta x)}\partial'_{\rho}\right)
     \hat x^{\mu} \Sigma_{\mu\nu},
\label{hhatx}
\end{eqnarray}
is evaluated as follows. 
By using 
the projection tensor defined by 
\begin{equation}
 P^{\mu\nu}=\eta^{\mu\nu}+\hat x^{\mu} \hat x^{\nu}, 
\end{equation} 
and with the aid of the relation 
$\partial'_{,\nu}\Sigma_{\mu}{}^{\nu}{}=0$, 
the second term in (\ref{hhatx}) becomes 
\begin{eqnarray}
&-&\int d^4 x'
   {F(\sigma,z)\over (\hat x\cdot \Delta x)}
   \hat x^{\rho}\partial'_{\rho}
     (\hat x^{\mu} \Sigma_{\mu\nu})\cr
 && = -\int d^4 x'{F(\sigma,z)\over (\hat x\cdot \Delta x)}
  \partial'_{\rho} \left( P^{\rho\mu}\Sigma_{\mu\nu}\right)\cr
 && = -\int d^4 x' {x'_{\rho} P^{\rho\mu}\Sigma_{\mu\nu}
    \over (\hat x\cdot \Delta x)^2}  \hat x^\eta \partial'_{\eta}
    F(\sigma,z)\cr
 && = -\int d^4 x'{F(\sigma,z)\over (\hat x\cdot \Delta x)^2}
\cr &&\qquad\times
    x'_\rho\left({2\over (\hat x\cdot \Delta x)}-\hat x^\eta \partial'_{\eta}\right)
     P^{\rho\mu}\Sigma_{\mu\nu}.
\end{eqnarray} 
In the second equality, we have used the relation 
\begin{eqnarray}
 \partial'_{\rho} f(\sigma)
 & = & {\Delta x_\rho \hat x^\eta\over (\hat x\cdot \Delta x)} 
\partial'_{\eta} f(\sigma),  
\end{eqnarray}
derived from (\ref{form1}) and (\ref{form2}).
Then we obtain
\begin{eqnarray}
\hat x^{\mu} h_{\mu\nu} & \sim & 
 \int d^4 x'{F(\sigma,z) \over (\hat x\cdot \Delta x)^2} 
  \Biggl[\hat x^{\mu}\Sigma_{\mu\nu}\cr
  &&- x'_\rho\left({2\over (\hat x\cdot \Delta x)}-\hat x^\eta \partial'_{\eta}\right)
     P^{\rho\mu}\Sigma_{\mu\nu}\Biggr]. 
\label{hform3}
\end{eqnarray}

The component further contracted with $\hat x^{\nu}$ is given by 
\begin{eqnarray}
&& \hat x^{\mu} \hat x^{\nu} h_{\mu\nu}  \sim  
 \int d^4 x'{F(\sigma,z) \over (\hat x\cdot \Delta x)^2} 
  \Biggl\{\hat x^{\mu}\hat x^{\nu}\Sigma_{\mu\nu}
\cr &&  
- {2\over (\hat x\cdot \Delta x)}x'_\rho 
   P^{\rho\mu}\hat x^{\nu}\Sigma_{\mu\nu}
    +\hat x^\eta x'_\rho \partial'_{\eta}
     (P^{\rho\mu}\hat x^{\nu}\Sigma_{\mu\nu})
\Biggr\}. 
\end{eqnarray}
By using the trace free condition $(P^{\mu\nu}-\hat x^\mu\hat x^\nu)
 \Sigma_{\mu\nu}=0$, 
the first and last terms in the round brackets 
are combined and simplified as 
\begin{eqnarray}
\int d^4 &x'&{F(\sigma,z) \over (\hat x\cdot \Delta x)^2} 
\left[\hat x^{\mu}\hat x^{\nu}\Sigma_{\mu\nu}+
    \hat x^\eta x'_\rho \partial'_{\eta}
     (P^{\rho\mu}\hat x^{\nu}\Sigma_{\mu\nu})\right]\cr
 & = &\int d^4 x'{F(\sigma,z) \over (\hat x\cdot \Delta x)^2} 
    \partial'_{\eta} (x'_\rho 
     P^{\rho\mu}P^{\eta\nu}\Sigma_{\mu\nu})\cr
 & = &\int d^4 x'{(x'_\rho P^{\rho\mu} x'_{\eta}P^{\eta\nu}
     \Sigma_{\mu\nu}) \over (\hat x\cdot \Delta x)^3} 
    \hat x^\xi \partial'_{\xi} F(\sigma,z) \cr
 & = & 
   -\int d^4 x'{F(\sigma,z) \over (\hat x\cdot \Delta x)^3} 
\cr &&\quad\times
    \left[{3\over (\hat x\cdot \Delta x)}-\hat
     x^{\eta}\partial'_{\eta} \right] (x'_\rho 
     P^{\rho\mu}x'_\eta P^{\eta\nu}\Sigma_{\mu\nu}). 
\end{eqnarray}
We finally arrive at the expression 
\begin{eqnarray}
\hat x^{\mu} \hat x^{\nu} h_{\mu\nu} & \sim & 
   -\int d^4 x'{F(\sigma,z) \over (\hat x\cdot \Delta x)^3} 
    \Biggl\{2 x'_\rho P^{\rho\mu}\hat x^{\nu}\Sigma_{\mu\nu}
\cr &&
    +\left[{3\over (\hat x\cdot \Delta x)}-\hat
     x^{\eta}\partial'_{\eta} \right] (x'_\rho 
     P^{\rho\mu}x'_\eta P^{\eta\nu}\Sigma_{\mu\nu})\Biggr\}. 
\label{hform4}
\end{eqnarray}
Hence, we find 
$$
\hat x^{\mu} h_{\mu\nu} = O\left({1\over\sigma^3}\right),
\quad 
\hat x^{\mu} \hat x^{\nu} h_{\mu\nu} = O\left({1\over\sigma^4}\right).
$$
in determining 
the order of $\sigma$ in the above expressions, 
we have used the assumption that support of $\Sigma_{\mu\nu}(x')$ 
is finite. 
Thus, $x'$ is supposed to be finite.  
As we have anticipated, the components of metric perturbations 
contracted with $\hat x^{\mu}$ have extra-suppression compared with 
the other components. 

The above result can be understood as a slight modification 
of the well-known fact that gravitational wave perturbations 
do not have components in the direction of propagation. 
Since the source of gravitational waves 
have extension in the present case, 
the components in $\hat x$-direction are
suppressed but they do not vanish completely. 

\section{Coordinate transform to regular coordinates}

In the preceding section, we have studied the asymptotic 
behavior of metric perturbations. From the results obtained 
in the preceding section, however, it is not clear  
whether the resulting metric is regular or not because 
the coordinates used there become singular when we take the limit 
of our interest. In this section, we perform coordinate 
transformation, and make it manifest that the metric 
obtained in the preceding section is regular. 

Using the Milne coordinates defined above, the bulk metric is written as 
\begin{equation}
  ds^2 = {\ell^2\over z^2}\left(dz^2 - dT^2+T^2 d\Omega^2 \right), 
\label{Tmetric}
\end{equation}
where $d\Omega^2$ is the squared line element of 
3-dimensional unit hyperboloid. 

Further we introduce new coordinates by 
\begin{equation}
 u=T-z, \quad \zeta=(T+z)^{-1}.
\end{equation}
In these coordinates the metric becomes 
\begin{equation}
 ds^2={\ell^2\over (1+u\zeta)^2}\left(
    du\, d\zeta+(1-u\zeta)^2 d\Omega^2 \right),  
\end{equation}
which is regular at $\zeta\to 0$
($\sigma \to \infty$) 
for finite $\sigma/z(\approx T/z\approx (1+u\zeta)/(1-u\zeta))$. 
Noting that 
\begin{equation}
 \sigma^2=T^2-2T \hat x^{\mu}(\Omega) x'_{\mu}+|x'|^2, 
\end{equation}
we find 
\begin{eqnarray}
 &&{1\over\sigma}=2\zeta(1+\zeta(-u+2 x'_{\mu}  
            \hat x^{\mu}(\Omega)+\cdots)),\cr
 &&{1\over (\hat x\cdot \Delta x)}=2\zeta(1+\zeta(-u+2 x'_{\mu}  
            \hat x^{\mu}(\Omega)+\cdots)),\cr
 && {1\over z} =2\zeta(1+u\zeta +\cdots),\cr
 && \sigma-z=u-x'_{\mu} \hat x^{\mu}(\Omega)+\cdots.
\label{sigmaz}
\end{eqnarray}

The components of metric perturbations in these regular coordinates 
$(u,\zeta,\Omega^{\alpha})$ 
are obtained from 
(\ref{hform2}), (\ref{hform3}), (\ref{hform4}), 
by using relations 
\begin{equation}
 {\partial x^{\mu}\over \partial \zeta}
   =-{\hat x^{\mu}(\Omega)\over 2\zeta^2}, 
\quad
 {\partial x^{\mu}\over \partial u}
   ={\hat x^{\mu}(\Omega)\over 2}, 
\quad
 {\partial x^{\mu}\over \partial\Omega^{\alpha}}
   ={1+u\zeta\over 2\zeta} {\partial\hat x^{\mu}
     \over \partial\Omega^{\alpha}}. 
\end{equation}
Then, substituting (\ref{sigmaz}), it will be easy to see that 
all components of metric perturbations in 
these regular coordinates are regular
at the future Cauchy horizon. 
(Thus any curvature invariants are also regular.) 
Though its importance is not apparent, it might be worth 
mentioning that $u$-components of metric perturbations are 
much more suppressed than the other components. 

\section{discussion}

In this paper we have studied the asymptotic 
behavior of metric perturbations 
near the future Cauchy horizon in the model 
of warped 5th dimension proposed by Randall and Sundrum. 
In the present analysis, we carefully took into account 
the contributions from all the K-K modes and the 
conservation of the energy-momentum tensor which 
becomes the source of metric perturbations.
As opposed to the claims in literature, our conclusion is 
that metric perturbations are regular at the 
future Cauchy horizon. 

Both results obtained in Ref.\cite{gt} and in the 
present paper indicate that the asymptotic behavior of 
metric perturbations in Randall-Sundrum model is 
much better than those expected from naive analysis,  
in which it was reported that the contracted Weyl curvature 
is expected to diverge at the future Cauchy horizon. 
As we have mentioned in Introduction, there has been 
a worry about this model raised by Randall and Sundrum 
themselves.   
If we take a naive picture based on the mode-by-mode analysis, 
the non-linear interaction between K-K modes 
seems to become stronger as we move far away from the brane. 
Therefore, there appears an inevitable divergence if we 
try to write down 4-dimensional effective action including 
interaction terms by integrating out the dependence on the 5-th
direction. 
Although the direct solution to this problem 
has not been obtained yet, 
the result proved in the present paper strongly 
suggests the possibility that the apparent pathological feature in 
higher order perturbation scheme in the Randall-Sundrum model 
at the classical level is not physical. 
Then, it will be an interesting and also a challenging problem 
to develop a formalism to handle higher order perturbations 
in this model. 
We would like to return to this issue in future publication. 

\vspace{5mm}
\centerline{\bf Acknowledgments}
 The author would like to thank J.~Garriga, A.~Ishibashi, H.~Kodama, M.~Sasaki
and T.~Shiromizu for their interesting comments. 
This work is supported by Monbusho Grant-in-Aid No. 1270154. 

\appendix
\section{Can we go through the future Cauchy horizon?}
In the original Randall-Sundrum model, the brane stays at 
a constant $y$. However, if we consider non-trivial evolution of 
energy density of matter fields on the brane, 
the brane can move in $y-$direction. 
Despite this possibility, one may think that observers on the 
brane cannot cross the future Cauchy horizon because the scale factor 
of our universe will collapse to zero at $y=\infty$ if we identify it 
with $e^{-|y|/\ell}$. 
We will show that this naive intuition is not correct and that 
the brane can go through the horizon of AdS space located at 
$y=\infty$.

We represent the motion of the brane by the function 
$z=z_b(T)$. Then, from (\ref{Tmetric}), the metric induced on the brane becomes
\begin{eqnarray}
ds_b^2 
 = d\tau^2+a(\tau)^2 d\Omega^2,
\end{eqnarray}
with 
\begin{eqnarray}
d\tau={\ell\sqrt{1-(d z_b /dT)^2}\over z_b} dT, 
a^2(\tau)={\ell^2 T^2(\tau)\over z_b^2(T(\tau))}, 
\end{eqnarray}
An interesting situation arises when we consider the case 
in which $dz_b/dT\to 1$ for $T\to \infty$. In this case, 
$T$ and $z_b$ goes to infinity within a finite value of the 
cosmological time $\tau$, while $a(\tau)$ stays finite. 
Hence, observers living on the brane can cross 
the future Cauchy horizon with a finite scale factor 
within a finite proper time. 
This means that the AdS Cauchy horizon is not necessarily  
the end of the brane-world.


\begin{thebibliography}{99}
\bibitem{gia} N. Arkani-Hamed, S. Dimopoulos, G. Dvali,
Phys.Lett. B429 (1998) 263-272, hep-ph/9803315; 
Phys.Rev. D59 (1999) 086004, hep-ph/9807344;
I. Antoniadis, N. Arkani-Hamed, S. Dimopoulos, G. Dvali,
Phys.Lett. B436 (1998) 257-263, hep-ph/9804398.
\bibitem{RS1} L.~Randall and R.~Sundrum, Phys. Rev. Lett.{\bf 83}, 
3370 (1999), hep-ph/9905221.
\bibitem{RS2} L.~Randall and R.~Sundrum, Phys. Rev. Lett.{\bf 83},
4690 (1999), hep-ph/9906064.
\bibitem{CG} A.~Chamblin and G.W.~Gibbons, 
hep-th/9909130.
\bibitem{CHR} A.~Chamblin, S.W.~Hawking and H.S.~Reall, 
Phys. Rev. {\bf D61}, 065007 (2000), 
hep-th/9909205.
\bibitem{SMS} T.~Shiromizu, K.~Maeda and M.~Sasaki, 
gr-qc/9910076.
\bibitem{gt} J. Garriga and T. Tanaka, 
Phys. Rev. Lett.{\bf 84}, 2778 (2000), 
hep-th/9911055.
\bibitem{EHM} 
R.~Emparan, G.T.~Horowitz and R.C.~Myers,
JHEP 0001, 007 (2000),  
hep-th/9911043.
\bibitem{CGRT}
C.~Csa\'ki, M.~Graesser, L.~Randall and J.~Terning, 
hep-ph/9911406. 
\bibitem{Rub}
C.~Charmousis, R.~Gregory and V.~Rubakov, 
hep-th/9912160.
\bibitem{SSM} M.~Sasaki, T.~Shiromizu and K.~Maeda,
hep-th/9912233. 
\bibitem{tm}
T.~Tanaka and X.~Montes,
to appear in Nucl. Phys. {\bf B}, 
hep-th/0001092.
\bibitem{Chiba}
T.~Chiba, gr-qc/0001029. 
\bibitem{GKR}
S.B.~Giddings, E.~Katz and L.~Randall, 
hep-th/0002091. 
 


\bibitem{Fla}
E.E.~Flanagan, S.-H.H.~Tve and I.~Wasserman, 
JHEP 0003, 023 (2000), 
hep-ph/9909373. 
\bibitem{Bin} 
P.~Bin\'etruy, C.~Deffayet, U.~Ellwanger and D.~Langlois, 
Phys. Lett. {\bf B477}, 285 (2000), 
hep-th/9910219.
\bibitem{mukoyama1}
Phys. Lett. {\bf B473}, 241 (2000), 
hep-th/9911165.
\bibitem{Ida}
D.~Ida, 
gr-qc/9912002. 
\bibitem{GS} 
J.~Garriga and M.~Sasaki, 
hep-th/9912118.
\bibitem{MSM}
S.~Mukohyama, T.~Shiromizu and K.~Maeda, 
hep-th/9912287. 
\bibitem{MWBH}
R.~Maartens, D.~Wands, B.A.~Bassett and I. Heard, 
hep-ph/9912464.
\bibitem{Mukoyama}
S.~Mukohyama, 
hep-th/0004067. 
\bibitem{KIS}
H.~Kodama, A.~Ishibashi and O.~Seto, 
hep-th/0004160. 
\bibitem{Maar}
R.~Maartens, 
hep-th/0004166.
\bibitem{lang}
D.~Langlois, 
hep-th/0005025. 
\bibitem{BDBL}
C.~van~de~Bruck, M.~Dorca, R.~Brandenberger and A.~Lukas, 
hep-th/0005032. 
\bibitem{KS}
K.~Koyama and J.~Soda, 
hep-th/0005239.

\end{thebibliography}
\end{document}